\documentclass{PoS}

\usepackage{here,epsfig,multicol}
\usepackage{amsmath,amssymb,bbm}

\PoS{PoS(HEP2005)262}

\def\utfit{{\bf{U}}\kern-.24em{\bf{T}}\kern-.21em{\it{fit}}\@}

\title{Upper Bounds on Rare K and B Decays from MFV}

\ShortTitle{Upper Bounds on Rare K and B Decays from MFV}

\author{
  \speaker{Thorsten Ewerth}\thanks{Based on collaboration with
    C. Bobeth, M. Bona, A. J. Buras, M. Pierini, L. Silvestrini and
    A. Weiler \cite{Bobeth:2005ck}.}\\
  Institute of Theoretical Physics, Sidlerstrasse 5, CH-3012 Bern,
  Switzerland\\
  E-mail: \email{tewerth@itp.unibe.ch}
}

\abstract{
  Recently, we studied the branching ratios of rare $K$ and $B$ decays
  in models with Minimal Flavor Violation (MFV) using the presently
  available information from the universal unitarity triangle analysis
  and from the measurements of ${\rm Br}(B\to X_s\gamma)$, ${\rm
  Br}(B\to X_s l^+l^-)$ and ${\rm Br}(K^+\to\pi^+\nu\bar\nu)$
  \cite{Bobeth:2005ck}.
  We analyzed in detail possible scenarios with positive or negative
  interference of Standard Model (SM) and New Physics contributions.
  In particular, we derived upper bounds on
  various rare decays and pointed out an interesting triple
  correlation between $B\to X_s\gamma$, $B\to X_s l^+l^-$ and
  $K^+\to\pi^+\nu\bar\nu$ present in MFV models.
}

\FullConference{
  International Europhysics Conference on High Energy Physics\\
  July 21st - 27th 2005\\
  Lisboa, Portugal
}

\begin{document}

%----- figure 1 -----
\begin{figure}[b]
\begin{center}
 \epsfig{figure=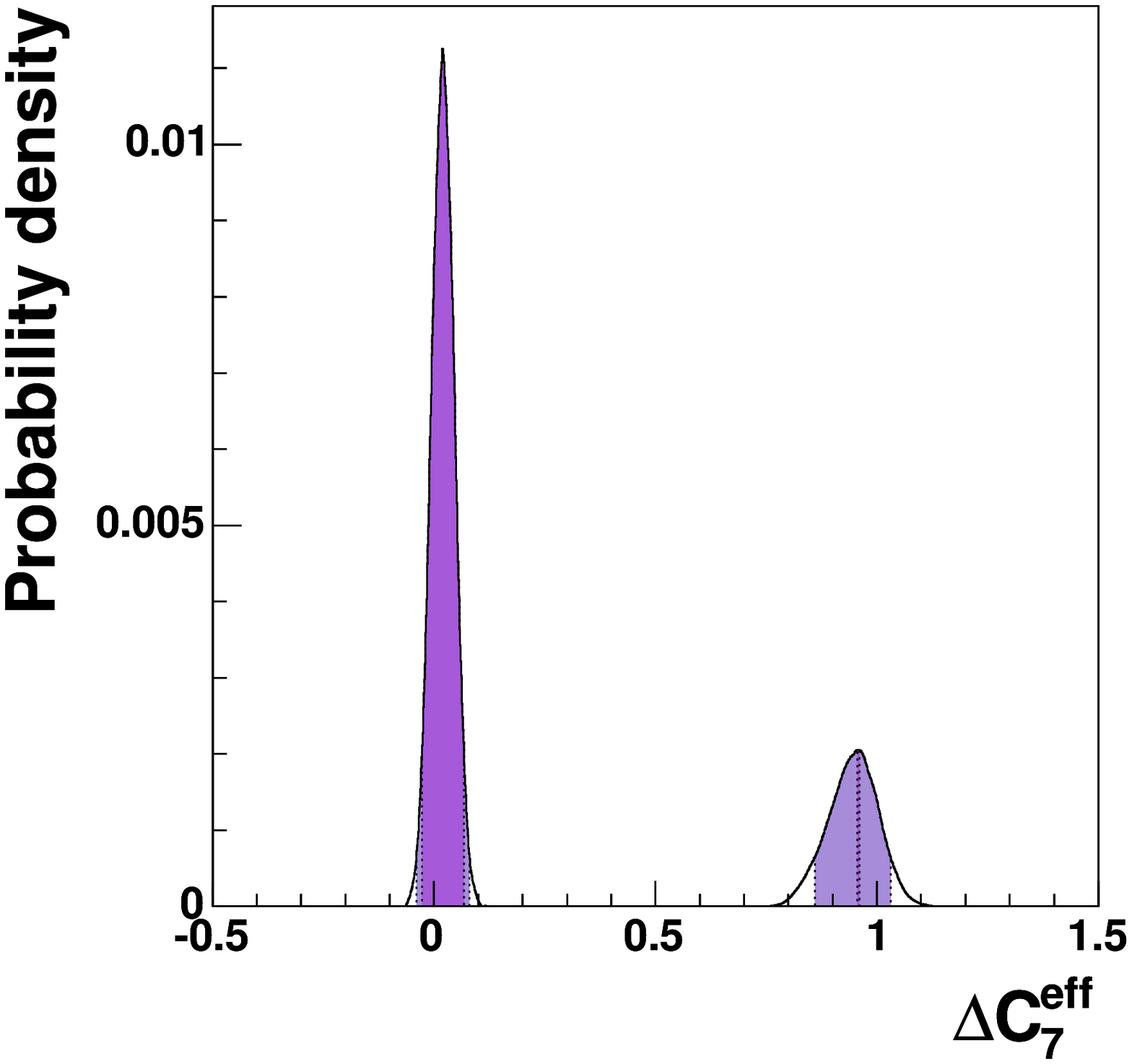,height=3.5cm}\hspace{.2cm}
 \epsfig{figure=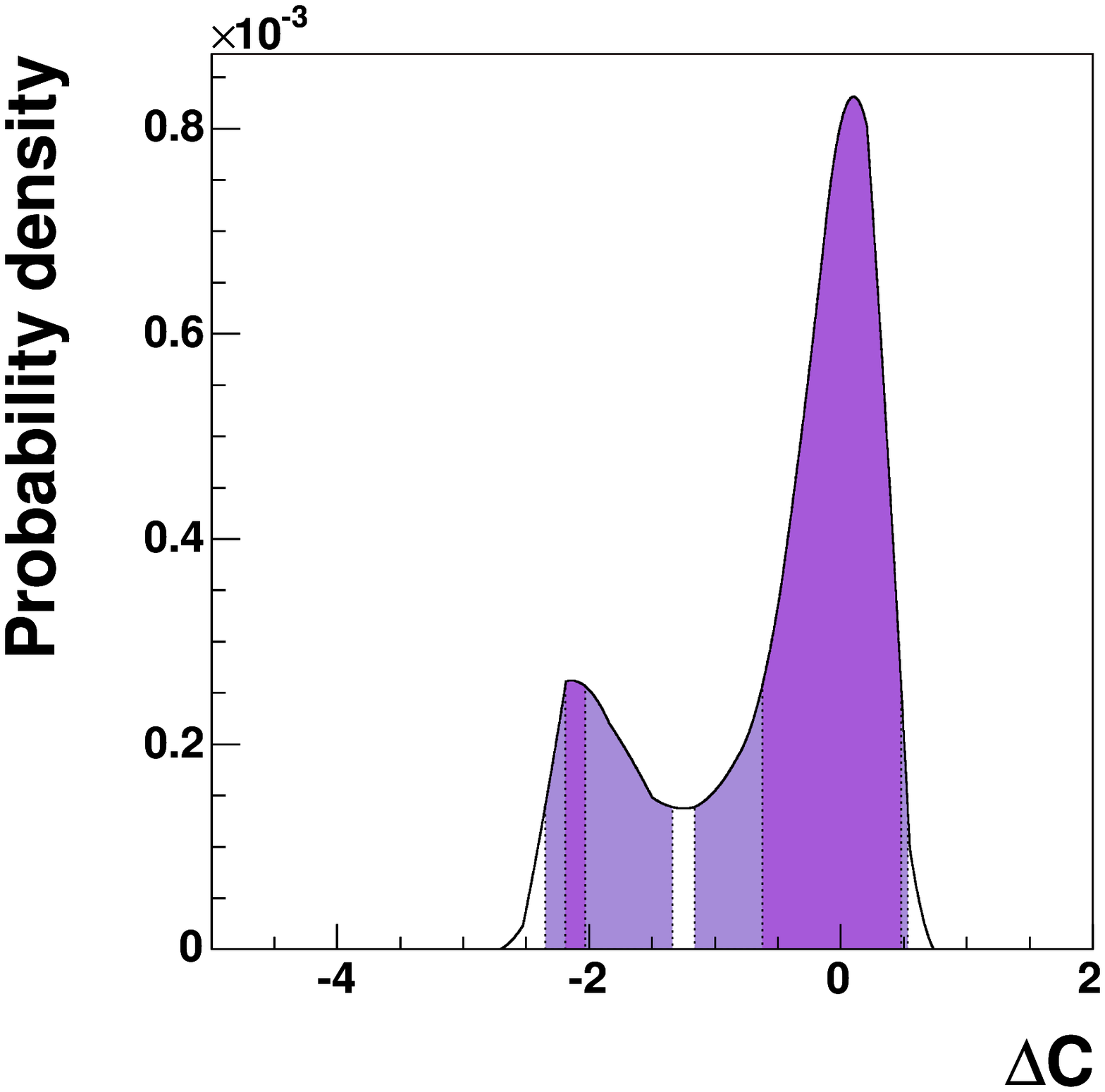,height=3.5cm}\hspace{.2cm}
 \epsfig{figure=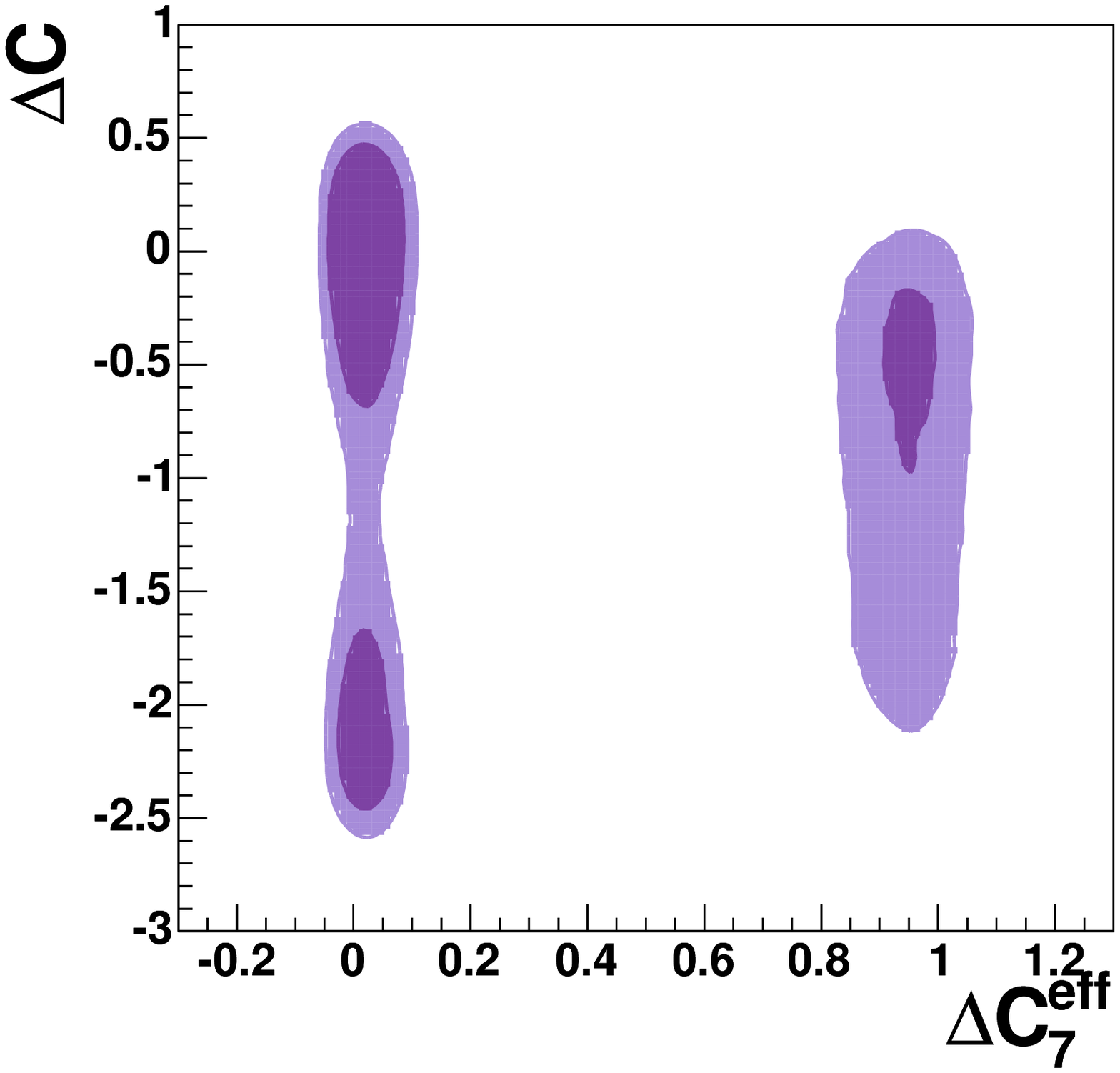,height=3.5cm}
\caption{\sl Probability density functions (pdfs) for $\Delta C_7^{\rm
  eff}$, $\Delta C$ and $\Delta C_7^{\rm eff}$ vs $\Delta C$. Dark
  (light) areas correspond to the 68\% (95\%) probability regions. The
  SM predictions are $C_7^{\rm eff}\approx -0.33$ and
  $C\approx 0.81$.}\label{figure1}
\end{center}
\end{figure}

The MFV models considered here are defined as in \cite{Buras:2000dm},
which means that {\it(i)} flavor and CP violation is entirely governed
by the CKM matrix, and {\it (ii)} the relevant operators entering the
effective Lagrangians of weak decays are those already present in the
SM \footnote{This is a special case of the effective field theoretical
approach given in \cite{D'Ambrosio:2002ex}.}. Under these assumptions
any weak decay amplitude can be written as follows \cite{Buras:2003jf},
{\small
\begin{equation*}\hspace{-4.8cm}
 A_{\rm decay} = {\sum}_i\,B_i\,\eta^i_{\rm QCD}\,
 V^i_{\rm CKM}\,F_i(v),\qquad F_i(v) = F^i_{\rm SM}+\Delta F_i(v)
\end{equation*}}
\!\!where the master functions $F_i(v)$, with $v$ denoting collectively
the parameters of a given MFV model, contain besides the heavy degrees
of freedom of the SM also the New Physics contributions via the real
functions $\Delta F_i(v)$. Furthermore, $B_i$ parameterize hadronic
matrix elements of local operators, and $\eta^i_{\rm QCD}$ describe
the QCD evolution of the functions $F_i(v)$ from the high- to the
low-energy scale. Both are universal in MFV and known from SM
calculations. For the CKM parameters $V^i_{\rm CKM}$ we use the
results of the universal unitary triangle analysis of the {\utfit}
collaboration \cite{Bona:2005eu}. In the following table we collect
the most important rare decays relevant for our numerical analysis
together with the master functions $F_i(v)$ characterizing them,
\begin{center}
{\small
\begin{tabular}{lcl}
  $K^+\to\pi^+\nu\bar\nu$,\;$K_L\to\pi^0\nu\bar\nu$,\;$B\to
   X_{d,s}\nu\bar\nu$ &\qquad& $X(v)=C(v)+B^{\nu\bar\nu}(v)$ \\[1mm]
  $K_{\rm L}\to\mu^+\mu^-$,\;$B_{d,s}\to l^+l^-$ &\qquad&
   $Y(v)=C(v)+B^{l\bar l}(v)$ \\[1mm]
  $B\to X_s\gamma$ &\qquad& $D'(v)$,\;$E'(v)$ \\[1mm]
  $B\to X_sl^+l^-$ &\qquad&
   $Y(v)$,\;$Z(v)=C(v)+\frac{1}{4}D(v)$,\;$E(v)$,\;$D'(v)$,\;$E'(v)$
   \\[1mm]
\end{tabular}}
\end{center}
Here, the master functions $E(v)$, $D(v)$ and $B(v)$ can be set to a
very good approximation equal to their SM values\footnote{In fact, varying
$\Delta D(v)$ in the range $\pm\,D_{\rm SM}$ has only little impact on
our numerical results. Furthermore, we checked that as long as the
functions $\Delta B^{l\bar l}(v)$ and $\Delta B^{\nu\bar\nu}(v)$ are
of comparable size as the SM ones our results do not change sizably.
Finally, $\Delta E(v)$ can be neglected due to a tiny pre-factor
accompanying it in the analytic expressions for the branching
ratios.}, and the functions $D'(v)$ and $E'(v)$ can be traded for
$C_7^{\rm eff}(v)$ being the relevant quantity at the low-energy scale
for both $B\to X_s\gamma$ and $B\to X_sl^+l^-$. Therefore the only
free parameters for the considered decays in the MVF models are the
functions $C_7^{\rm eff}(v)$ and $C(v)$. The strategy for our
numerical analysis is then {\it(i)} to constrain the functions
$C_7^{\rm eff}(v)$ and $C(v)$ through the measured branching ratios of
$B\to X_s\gamma$, $B\to X_sl^+l^-$ and $K^+\to\pi^+\nu\bar\nu$, and
{\it(ii)} to compute expectation values and upper bounds following
from the constraints on these two parameters for the other decays of
the above given table.

%----- figure 2 -----
\begin{figure}[t]
\begin{center}
 \epsfig{figure=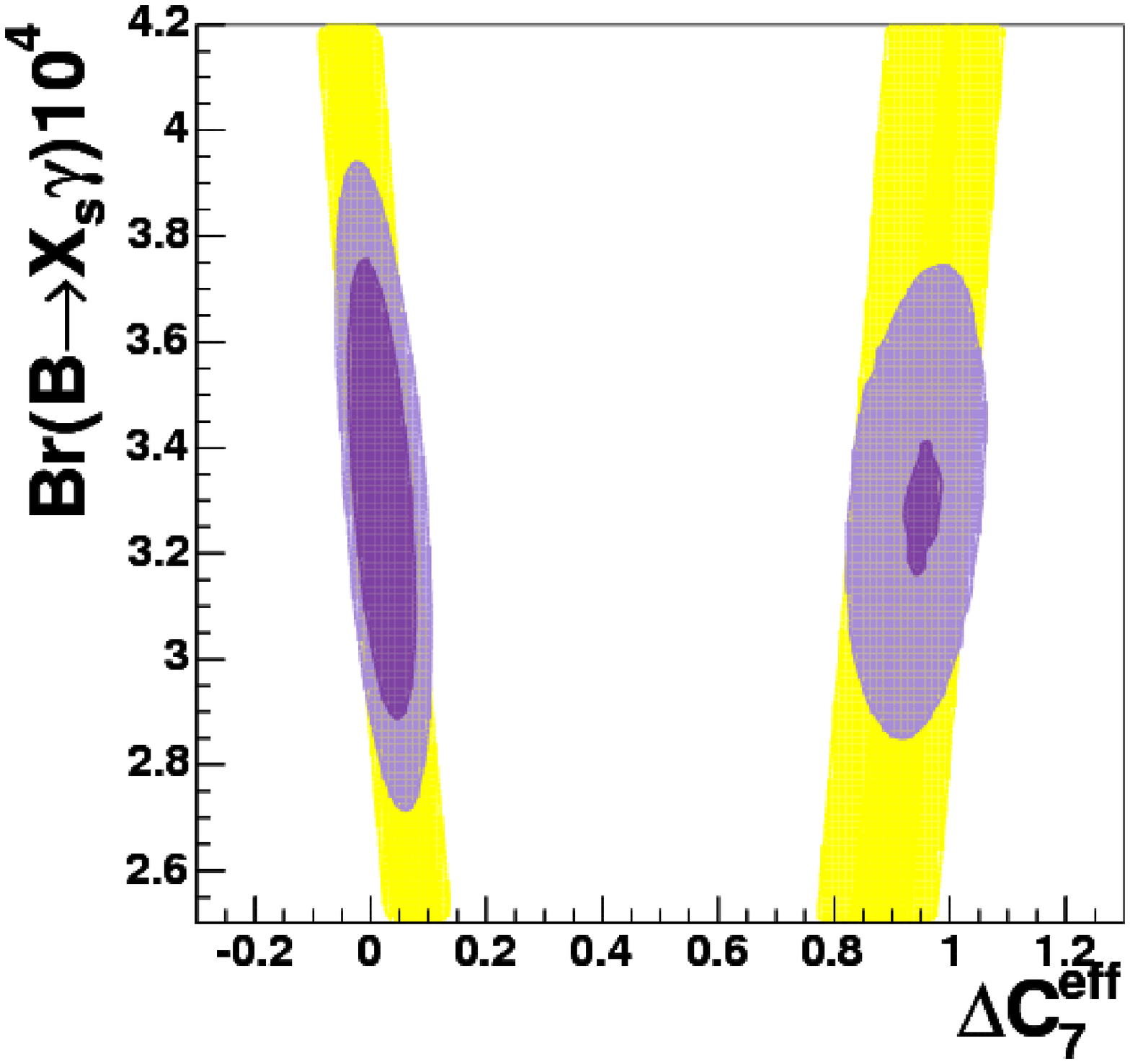,height=3.5cm}\hspace{.4cm}
 \epsfig{figure=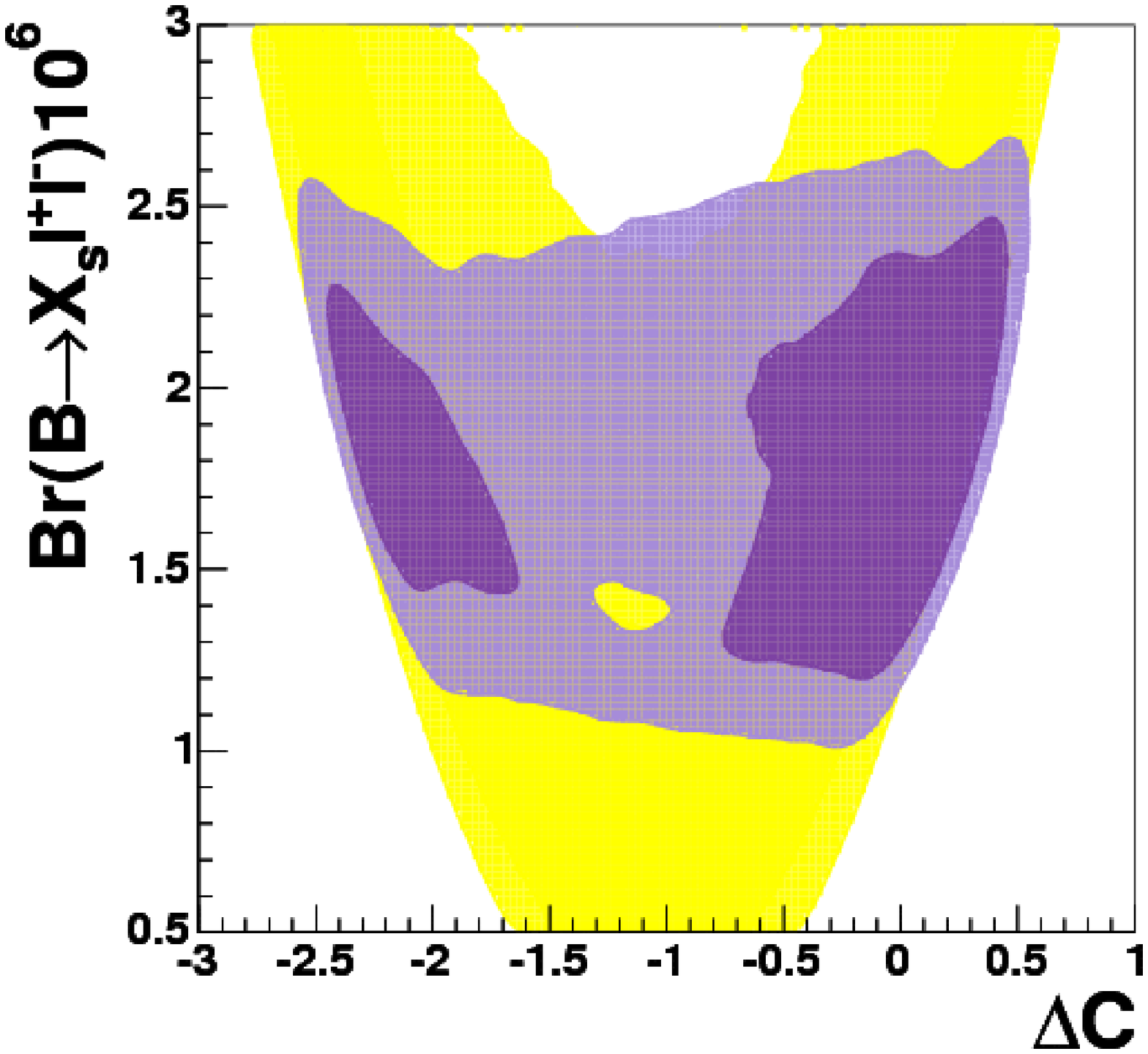,height=3.5cm}\hspace{.4cm}
 \epsfig{figure=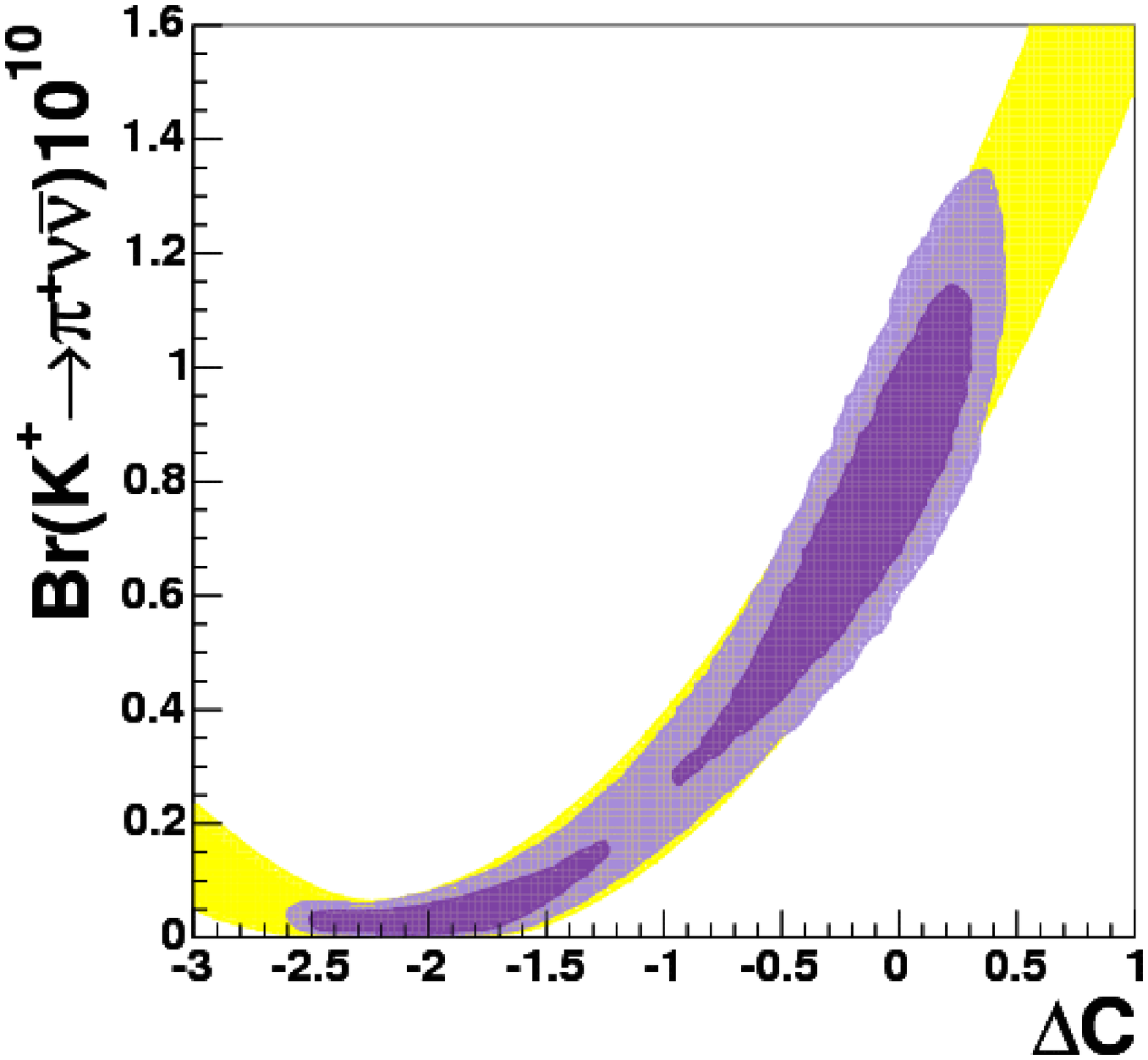,height=3.5cm}
\caption{\sl Pdfs for the branching ratios used to constrain $C_7^{\rm
  eff}$ and $C$ as functions of these parameters. The SM predictions
  are ${\rm Br}(B\to X_s\gamma)_{E_\gamma\,>\,1.8\,{\rm
  GeV}}\approx3.5\times10^{-4}$, ${\rm Br}(B\to
  X_sl^+l^-)_{1\,<\,q^2\,({\rm GeV})\,<\,6}\approx1.6\times10^{-6}$
  and ${\rm Br}(K^+\to\pi^+\nu\bar\nu)\approx8.3\times10^{-10}$. Dark
  (light) areas correspond to the 68\% (95\%) probability regions, and
  very light ones to the range without using the experimental
  information.}\label{figure2}
\end{center}
\end{figure}

%----- figure 3 -----
\begin{figure}[b]
\begin{center}
 \epsfig{figure=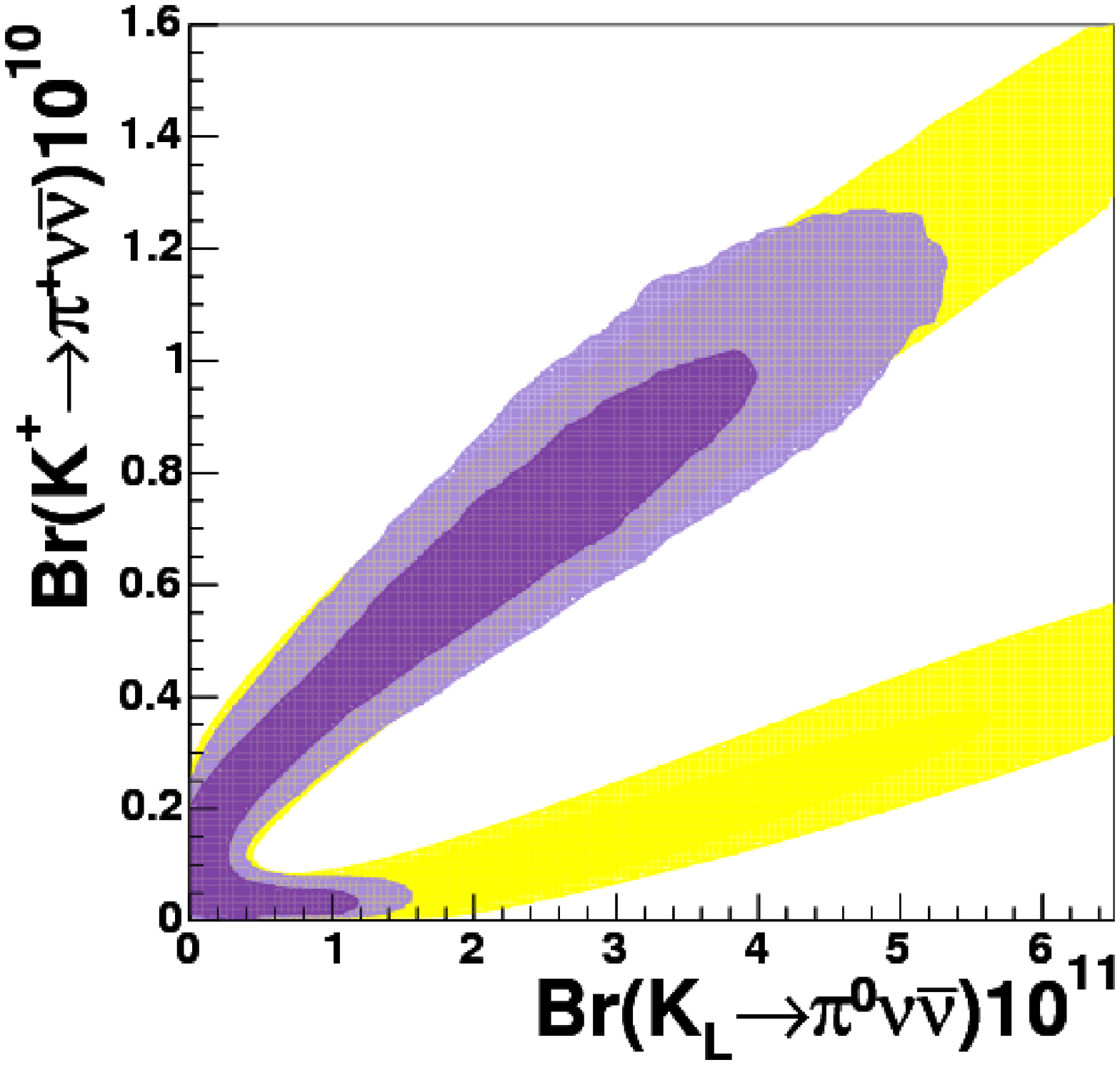,height=3.5cm}\hspace{.6cm}
 \raisebox{2.5mm}{\epsfig{figure=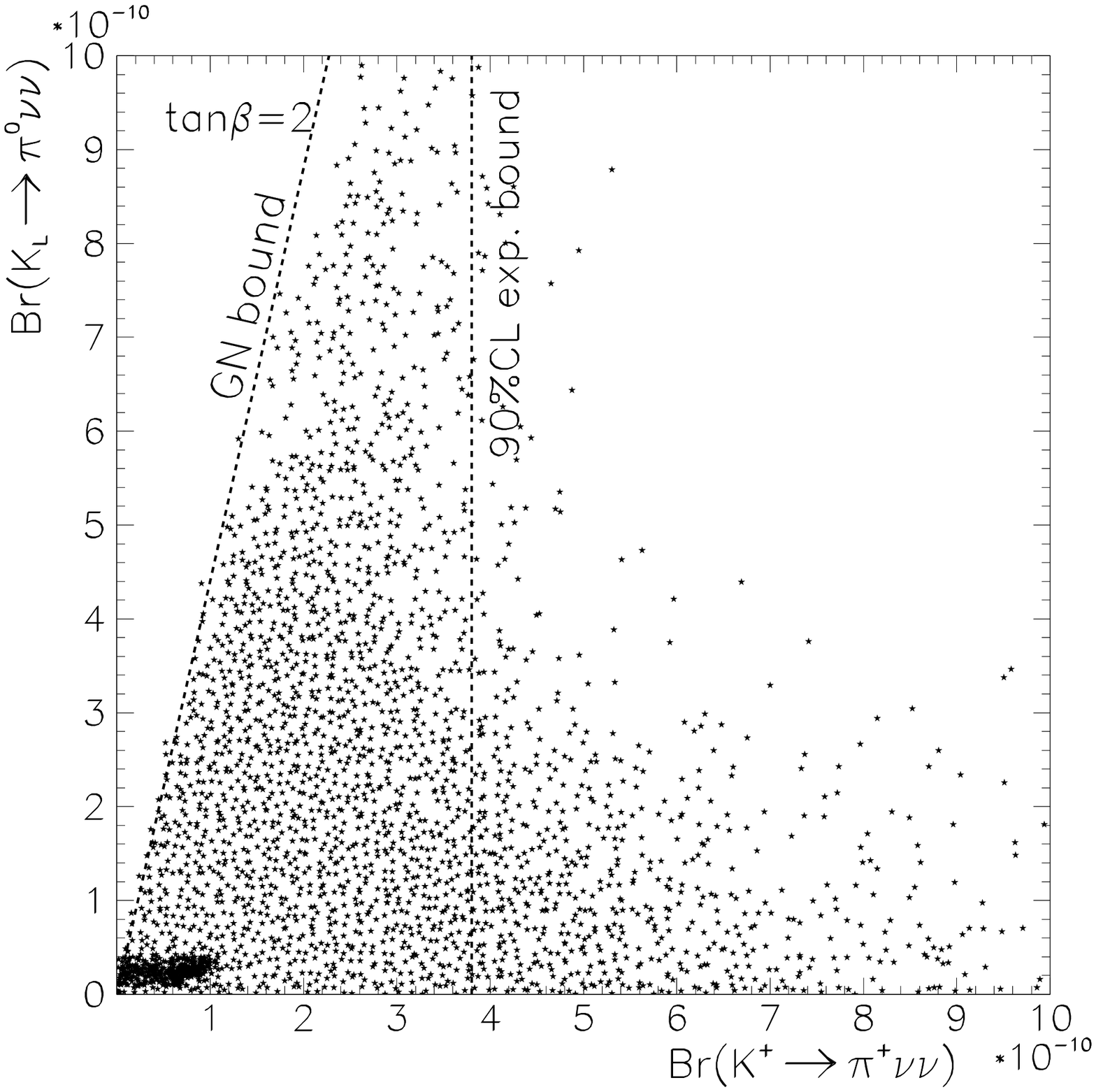,height=3.1cm}}
\caption{\sl Left: Pdf for ${\rm Br}(K^+\to\pi^+\nu\bar\nu)$ vs ${\rm
  Br}(K_L\to\pi^0\nu\bar\nu)$ within MFV models. Dark (light) areas
  correspond to the 68\% (95\%) probability regions, and very light
  ones to the range without using the experimental information.
  Right: Scatterplot for ${\rm Br}(K^+\to\pi^+\nu\bar\nu)$ vs ${\rm
  Br}(K_L\to\pi^0\nu\bar\nu)$ within the general MSSM.}\label{figure3}
\end{center}
\end{figure}

\begin{figure}[t]
\begin{center}
 \epsfig{figure=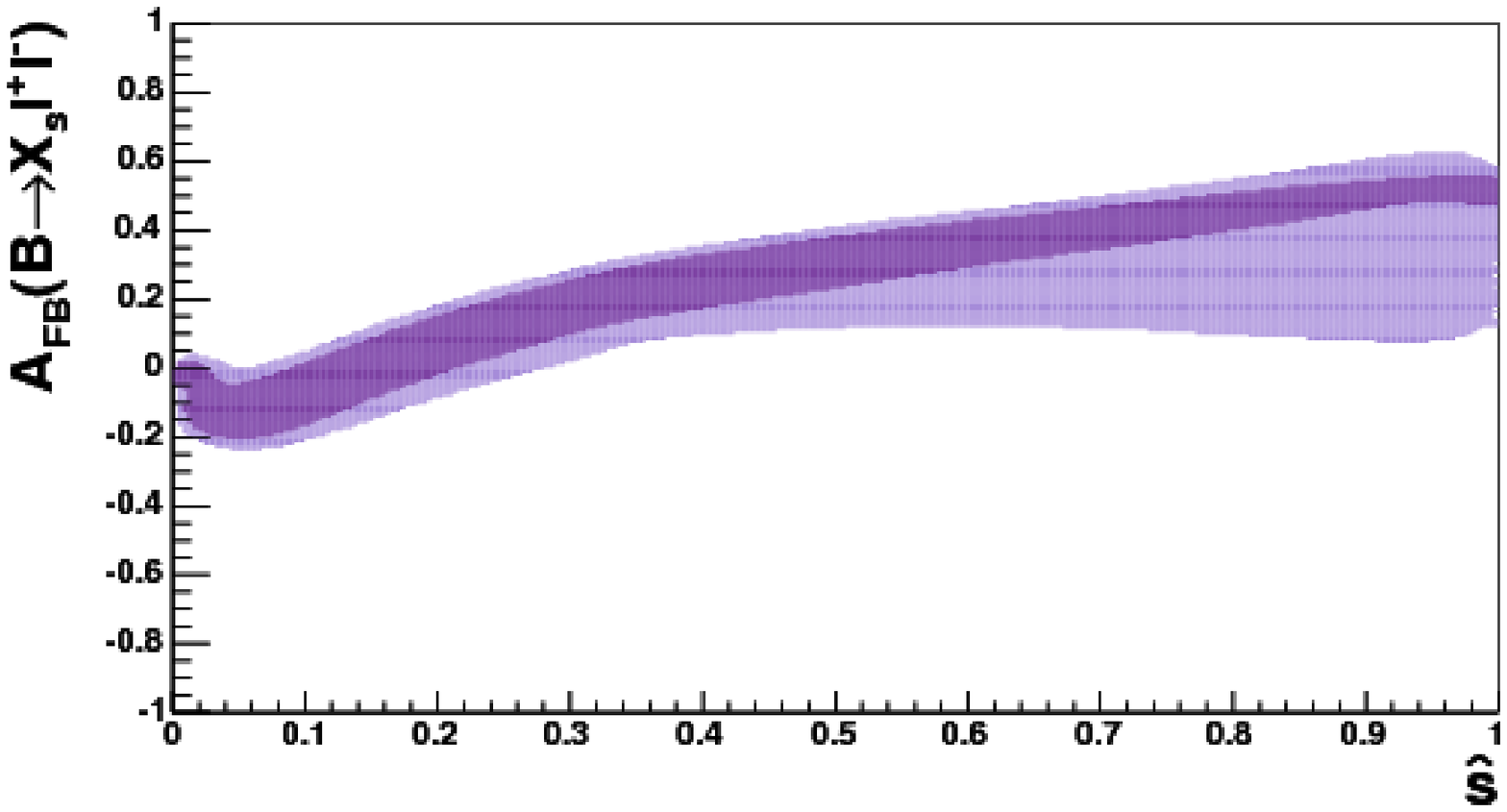,height=3cm}\hspace{.2cm}
 \epsfig{figure=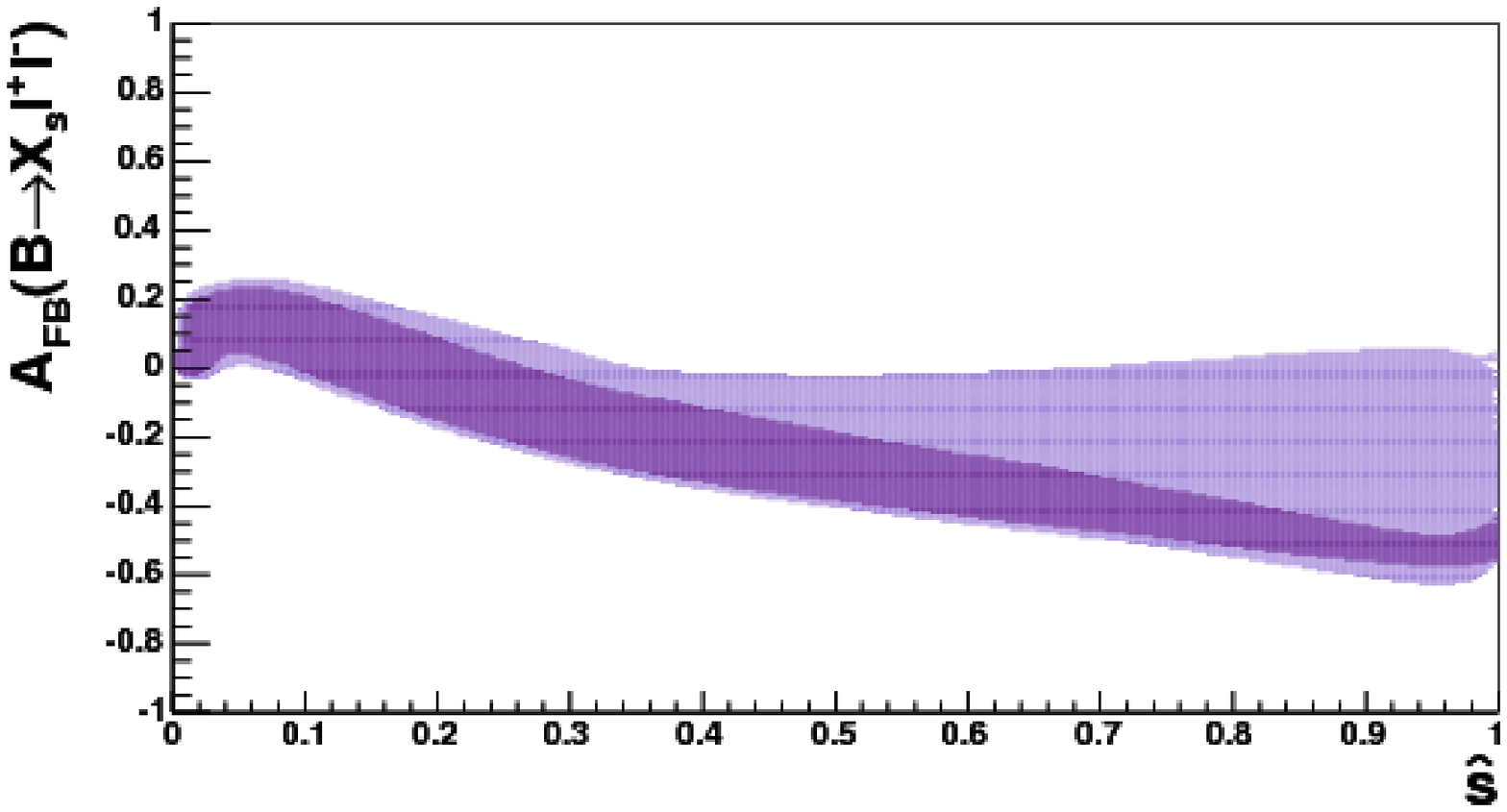,height=3cm}
\caption{\sl Forward-backward asymmetry of $B\to X_sl^+l^-$ for the
  SM-like solution for $C_7^{\rm eff}$ with $\Delta C>-1$ (left) and
  $\Delta C<-1$ (right). Dark (light) areas correspond to the 68\%
  (95\%) probability regions.}\label{figure4}
\end{center}
\end{figure}

Figure \ref{figure1} shows that there exist SM-like solutions with
$C_7^{\rm eff}(v)<0$ and $C(v)>0$. But we also find another solution
for $C_7^{\rm eff}(v)$ which corresponds to reversing the sign of
the SM prediction. This opposite sign solution is however disfavored
because ${\rm Br}(B\to X_sl^+l^-)$ tends to become larger than the
experimental value \cite{Gambino:2004mv}. Furthermore, we find a
second solution for $C(v)$ which also corresponds to reversing the
sign of its SM prediction. The suppression of this opposite sign
solution follows from the experimental data on ${\rm
Br}(K^+\to\pi^+\nu\bar
\nu)$. The third plot shows the correlation between $\Delta C_7^{\rm
eff}(v)$ and $\Delta C(v)$. There are two solutions for $\Delta C(v)$
for the SM-like solution of $\Delta C_7^{\rm eff}(v)$ but only one
solution for $\Delta C(v)$ for the opposite sign solution of $C_7^{\rm
eff}(v)$. The first plot in Figure \ref{figure2} gives the pdf for
${\rm Br}(B\to X_s\gamma)$ vs $\Delta C_7^{\rm eff}(v)$ with the
running charm quark mass used in the low-energy matrix elements
and a photon cut-off $E_\gamma=1.8$ GeV. Again we see that the
opposite sign solution is disfavored. The second plot shows the pdf
for ${\rm Br}(B\to X_sl^+l^-)$ vs $\Delta C(v)$ in the low-$q^2$
region in order to avoid the theoretical uncertainties due to
$c{\bar c}$ resonances, and the last one ${\rm
Br}(K^+\to\pi^+\nu\bar\nu)$ vs
$\Delta C(v)$. As can be seen, the signs of $C(v)$ and $C_7^{\rm
eff}(v)$ cannot be determined at present, but with the ongoing
reduction of the experimental and theoretical uncertainties of rare
$B$ decays it will become possible to eliminate the opposite sign
solution of $C_7^{\rm eff}(v)$. However, it will be difficult to
determine the sign of $C(v)$ from ${\rm Br}(B\to X_sl^+l^-)$ alone.
This ambiguity can then be resolved by a more precise measurement of
${\rm Br}(K^+\to\pi^+\nu\bar\nu)$. We also note that eliminating the
opposite sign solution of $C(v)$ by means of this decay would
basically also eliminate the opposite sign solution of $C_7^{\rm
eff}(v)$. Figure \ref{figure3} shows ${\rm Br}(K^+\to\pi^+\nu\bar\nu)$
vs ${\rm Br}(K_L\to\pi^0\nu\bar\nu)$. Within the MFV models (left
plot) one has a very strong correlation and the neutral decay mode is
always one order of magnitude smaller than the charged one. The right
plot illustrates that this is completely different in non-MFV models.
It has been obtained by taking the full flavor structure of the squark
sector of the general MSSM into account \cite{Buras:2004qb}. Here, the
neutral decay mode can easily be of the same order as the charged one.
Other interesting sensitive probes of $C_7^{\rm eff}(v)$ and $C(v)$
are the forward-backward asymmetries of inclusive and exclusive
semileptonic $B$ decays. Figure \ref{figure4} shows the shape of the
asymmetry for $B\to X_sl^+l^-$ depending on the sign of $C(v)$ for the
SM-like solution of $C_7^{\rm eff}(v)$. A measurement of this
observable would clearly help in determining the sign of $C(v)$.
The table given below displays our upper bounds for rare decays in MFV
at 95\% probability together with the corresponding SM predictions and
the available experimental information.
\begin{center}
{\small
\begin{tabular}{|c|c|c|c|c|}\hline
 {Branching Ratios} & MFV (95\%) &  SM (95\%) &  SM (68\%) & exp. \\
 \hline
 ${\rm Br}(K^+\to\pi^+\nu\bar\nu)\times 10^{11}$ & $< 11.9$ & $< 10.9$
  & $8.3\pm 1.2$ & $(14.7^{+13.0}_{-8.9})$ \\ \hline
 ${\rm Br}(K_{\rm L}\to\pi^0\nu\bar\nu)\times 10^{11}$  & $< 4.6$
  & $<4.3$ & $3.1\pm 0.6$ & $ < 2.86\times10^{4}$ \\ \hline
 ${\rm Br}(K_{\rm L}\to\mu^+\mu^-)_{\rm SD}\times 10^{9} $ & $< 1.4$ &
  $< 1.2$ & $0.9\pm 0.1$ & - \\ \hline
 ${\rm Br}(B\to X_s\nu\bar\nu)\times 10^{5}$ & $< 5.2$ & $< 4.1$ &
  $3.7\pm 0.2$ & $< 64 $ \\ \hline
 ${\rm Br}(B\to X_d\nu\bar\nu)\times 10^{6}$ & $< 2.2$ & $< 1.9$ &
  $1.5\pm 0.2$ & - \\ \hline
 ${\rm Br}(B_s\to\mu^+\mu^-)\times 10^{9}$ &  $< 7.4$ & $< 5.9$ &
  $3.7\pm 1.0$ & $<2.7\times 10^{2}$ \\ \hline
 ${\rm Br}(B_d\to\mu^+\mu^-)\times 10^{10}$ &  $< 2.2$ & $< 1.8$ &
  $1.0\pm 0.3$ & $<1.5\times 10^3$ \\ \hline
\end{tabular}}
\end{center}
We conclude that the present constraints from $B\to X_s\gamma$, $B\to
X_sl^+l^-$ and $K^+\to\pi^+\nu\bar\nu$ do not allow for large
enhancements of the branching ratios of other rare $K$ and $B$ decays
with respect to the SM predictions. Any violation of these upper
bounds signals new sources of flavor and CP violation and/or the
presence of operators beyond those relevant in the SM.

%----- figure 4 -----

%%%

%\begin{multicols}{2}

%\end{multicols}

\end{document}